\documentclass[preprint,preprintnumbers,amsmath,amssymb]{revtex4}
\usepackage{graphicx}% Include figure files
\usepackage{dcolumn}% Align table columns on decimal point
\usepackage{bm}% bold math
\usepackage{epsfig}

\begin{document}

%\preprint{APS/123-QED}

\title{Critical surfaces for general inhomogeneous bond percolation problems}% Force line breaks with \\
\author{Christian R. Scullard}
\email{scullard@uchicago.edu}
\affiliation{Department of Geophysical Sciences, University of Chicago, Chicago, Illinois 60637, USA}
\author{Robert M. Ziff}
\email{rziff@umich.edu}
\affiliation{Michigan Center for Theoretical Physics and Department of Chemical Engineering, University of Michigan, Ann Arbor, Michigan 48109-2136, USA}
\date{\today}

\begin{abstract}
We present a method of general applicability for finding exact or accurate approximations to bond percolation thresholds for a wide class of lattices. To every lattice we sytematically associate a polynomial, the root of which in $[0,1]$ is the conjectured critical point. The method makes the correct prediction for every exactly solved problem, and comparison with numerical results shows that it is very close, but not exact, for many others. We focus primarily on the Archimedean lattices, in which all vertices are equivalent, but this restriction is not crucial. Some results we find are kagome: $p_c=0.524430...$, $(3,12^2): p_c=0.740423...$, $(3^3,4^2): p_c=0.419615...$, $(3,4,6,4):p_c=0.524821...$, $(4,8^2):p_c=0.676835...$, $(3^2,4,3,4)$: $p_c=0.414120...$ .
The results are generally within $10^{-5}$ of numerical estimates. For the inhomogeneous checkerboard and bowtie lattices, errors in the formulas (if they are not exact) are less than $10^{-6}$.\end{abstract}

\maketitle
%\pacs{Ak 64.60}% PACS, the Physics and Astronomy
                             % Classification Scheme.
%\keywords{Suggested keywords}%Use showkeys class option if keyword
                              %display desired
%\maketitle

\section{Introduction}
Since its introduction in the 1950s by Broadbent and Hammersley \cite{Hammersley,BroadbentHammersley}, percolation has been a rich source of interesting and challenging problems. Its very simple construction has made it the classic model of disordered media and phase transitions. However, despite its apparent simplicity, exact answers to even the most basic questions have been difficult to come by.

The bond percolation process is defined on a lattice by declaring each edge to be open with probability $p$ and closed with probability $1-p$, where $p \in [0,1]$. One can then ask many questions about the resulting connected bonds, such as the average size of open clusters, the cluster density, and the probability of crossing large regions. We are concerned here with the critical threshold, $p_c$, defined as the probability at which an infinite connected cluster first appears. Percolation thresholds depend on the lattice under consideration, with each one presenting a different set of challenges. However, there is a limited number of cases in which the critical probability is known exactly, with all currently solved problems confined to a particular self-dual class of lattices, as discussed below. Although we will be concerned mainly with the bond problem here, we could also consider site percolation in which $p$ is the occupation probability of the vertices (sites) on the graph. There are likewise very few exactly known site thresholds.

In the absence of exact solutions, we must seek alternative approaches. One is to study the problem numerically, and thresholds, both site and bond, are now known to high precision for a wide variety of systems \cite{SudingZiff99,Parviainen,Feng,ZiffGu,Becker,Quintanilla,Lee,Grassberger,Kownacki,Corsi,Kozakova2008,LiZhang09,BaekMinnhagenKim09,Ding}. Mathematically rigorous results in the form of inequalities have also been found \cite{Wierman2002b,Wierman2003,MayWierman05}, and the bounds are continually being narrowed. Riordan and Walters \cite{Riordan} have developed rigorous ``$99.9999\%$ confidence intervals,'' subtly distinct from rigorous bounds, in which simulations are used to place percolation thresholds in a certain range with very high probability. Another avenue of pursuit that has emerged is the search for ``universal'' formulas; closed expressions for the percolation threshold as a function of a few lattice parameters, usually focussing on coordination number and dimension (see \cite{Galam} and \cite{Wierman2005} for just two examples). These formulas generally meet with mixed success, providing good approximations for some lattices but performing poorly for others. Whether or not one believes a simple closed formula is a reasonable goal, it is clear that the next best thing is to have a well-defined procedure for finding a good analytical approximation to the threshold of any given lattice. We present such a procedure here for bond percolation. With each lattice we show how to associate a polynomial, the solution of which in $[0,1]$ is the prediction for the critical point. Some of the main results of this work were reported in a recent Letter \cite{ScullardZiff08}. Our approximation is related to Wu's homogeneity assumption \cite{Wu79} for the Potts model as both methods give similar results when applied to the same lattices. Wu's work goes back several decades, but recently he has applied it to a variety of new lattice systems \cite{Wu09}, though not the ones presented here.

\begin{center}
\begin{figure}
 \includegraphics{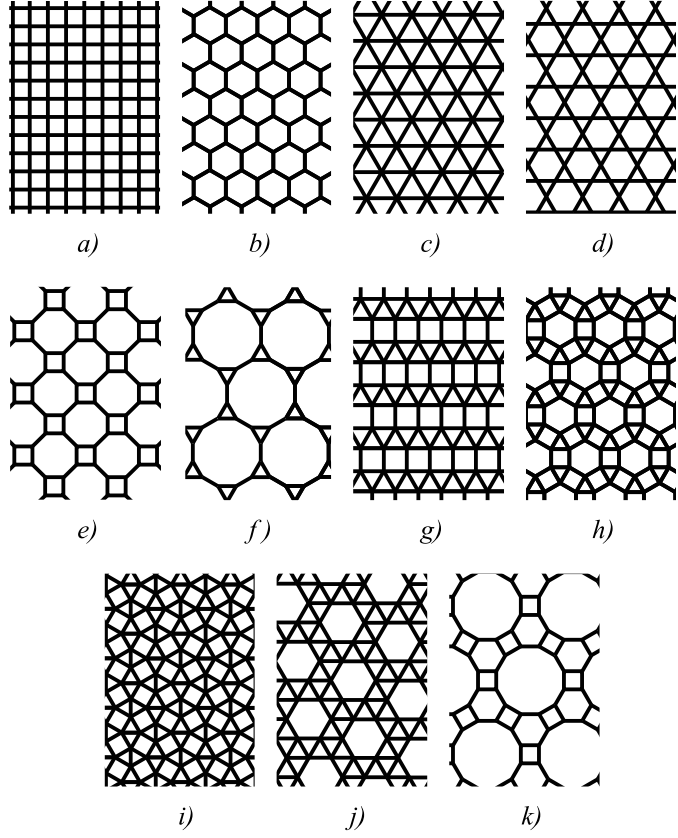} 
\caption{The Archimedean lattices; a) square; b) honeycomb; c) triangular; d) kagome; e) $(4,8^2)$; f) $(3,12^2)$; g) $(3^3,4^2)$; h) $(3,4,6,4)$; i) $(3^2,4,3,4)$; j) $(3^4,6)$; k) $(4,6,12)$ .} \label{fig:archimedean}
\end{figure}
\end{center}

Although our method works for any two-dimensional graph with a periodic structure, we confine our attention here to the Archimedean lattices, which are composed of regular polygons such that all vertices are equivalent. There are 11 such graphs and they are shown in Figure \ref{fig:archimedean}, where we have used the Gr\"unbaum and Shephard \cite{Grunbaum} notation: given any vertex, one lists, in order, the number of sides of the polygons adjacent to it. For example, the kagome lattice (Figure \ref{fig:archimedean}d), is $(3,6,3,6)$. If the same number would appear more than once in succession, an exponent is used, so the lattice in Figure \ref{fig:archimedean}g is named $(3^3,4^2)$ rather than $(3,3,3,4,4)$.
\begin{center}
\begin{figure}
 \includegraphics{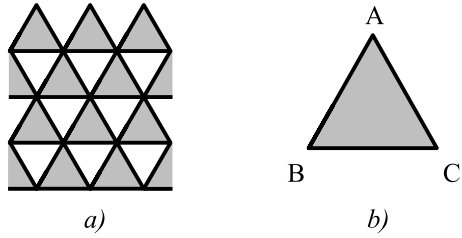} 
\caption{a) A class of exactly solvable lattices; b) every shaded triangle of a) can represent any network of sites and bonds contained in the vertices $(A,B,C)$.} \label{fig:selfdual}
\end{figure}
\end{center}
\section{Method}
Currently, the most general tool available for deriving exact percolation thresholds is the triangle-triangle transformation \cite{Ziff06}, which is based upon duality and is a generalization of the star-triangle transformation \cite{SykesEssam}. The method works only for lattices in which the basic cell is contained within three sites and arranged in a self-dual way \cite{ZiffScullard06}. The simplest example of a self-dual array of
triangles is just the triangular array itself, as shown in Figure \ref{fig:selfdual}. Solutions to these problems are derived by considering probabilities of events that take place on a single cell of the given lattice. For a lattice of the type in Figure \ref{fig:selfdual}, the critical threshold is identified using the condition given by \cite{Ziff06,ChayesLei}:
\begin{equation}
P(A,B,C)=P(\bar{A},\bar{B},\bar{C}) \label{eq:critcond}
\end{equation}
where $P(A,B,C)$ is the probability that the three vertices can be connected through open bonds in the cell, $P(\bar{A},\bar{B},\bar{C})$ is the probability that none are connected, and the shaded triangle can represent any network of sites and bonds. This condition is also related to Wu's criticality condition $qA = C$ (where $A$ and $C$ are elements of the Boltzmann factor of a single triangle) for the $q-$state Potts model on triangular arrays \cite{Wu06}.

\begin{center}
\begin{figure}
 \includegraphics{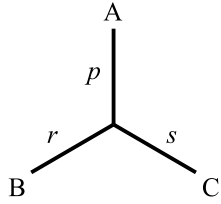} 
\caption{Using this simple star in Figure \ref{fig:selfdual}b results in the honeycomb lattice (Figure \ref{fig:archimedean}b).} \label{fig:star}
\end{figure}
\end{center}

If the cell is a simple star (Figure \ref{fig:star}) where each bond is given different probabilities, $p$,$r$,$s$, then (\ref{eq:critcond}) gives the criticality condition
\begin{equation}
H(p,r,s) \equiv prs - rp - rs - ps + 1=0 \label{eq:hex}
\end{equation}
which is the inhomogeneous threshold for the honeycomb lattice. By assigning a different probability to each bond of the unit cell, we have obtained a critical surface. We can find the homogeneous critical probability, $p_c$, by setting all probabilities equal, giving
\begin{equation}
H(p,p,p)= p^3-3 p^2+1=0
\end{equation}
so $p_c=1-2 \sin \pi/18 \approx 0.652704$ \cite{SykesEssam}.

\begin{center}
\begin{figure}
 \includegraphics{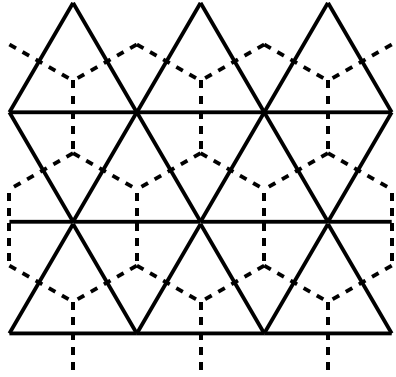} 
\caption{The duality transformation. The honeycomb and triangular lattices are dual pairs.} \label{fig:duality}
\end{figure}
\end{center}

A useful property of two-dimensional bond thresholds is the duality relationship. For a given graph, $L$, its dual, $L_d$, is constructed as shown in Figure \ref{fig:duality}; a vertex is placed in every face of $L$ and connected by bonds to all the vertices in neighbouring faces. Bond thresholds of a dual pair of lattices are related by \cite{SykesEssam}
\begin{equation}
 p_c(L)=1-p_c(L_d) . \label{eq:duality}
\end{equation}
This can be generalized, for lattices obeying (\ref{eq:critcond}), to inhomogeneous thresholds. The dual of the honeycomb lattice is the triangular lattice, so we can write
\begin{equation}
\mathrm{T}(p,r,s) \equiv -\mathrm{H}(1-p,1-r,1-s) \label{eq:inhomduality}
\end{equation}
giving 
\begin{equation}
\mathrm{T}(p,r,s) = p r s -p-r-s+1=0 \label{eq:triangular}
\end{equation}
which is the result we would derive using (\ref{eq:critcond}) directly. The homogeneous polynomial is $p^3-3 p+1=0$ with solution $p_c=2 \sin \pi/18 \approx 0.347296$ consistent with (\ref{eq:duality}). The minus sign in (\ref{eq:inhomduality}) is arbitrary given that we set these functions equal to $0$, but it ensures that the constant term in the critical function is $+1$, a convention we will always employ.
It is clear that any inhomogeneous problem, with probabilities $(p_1,p_2,...,p_n)$ on the bonds of the unit cell, that can be solved with (\ref{eq:critcond}) has a threshold of the form
\begin{equation}
 f(p_1,p_2,...,p_n)=0
\end{equation}
where no probability appears as a power greater than $1$, a property we will refer to as ``linearity''. The central assumption of the method presented here is that {\it all} bond percolation thresholds have this form. The goal is then to find the function $f$ for each lattice. This will be done by demanding that the function reduce correctly to all known special cases. For example, in the honeycomb lattice case, setting $p=1$ contracts the $p$ bond to zero length resulting in a square lattice. Equation (\ref{eq:hex}), predicts $\mathrm{S}(r,s) \equiv \mathrm{H}(1,r,s)=1-r-s=0$, which is the correct threshold \cite{SykesEssam}. The challenge is to find sufficient conditions for each $f$ to completely constrain the function.

One might worry that for lattices that are not handled by (\ref{eq:critcond}), the requirements of linearity of $f$ in all its probabilities and its correct reduction in all special cases might sometimes be in conflict. Interestingly, this does not appear ever to be the case, i.e., such an $f$ always seems to exist. Another potential difficulty is that there may be more than one $f$ that satisfies all the conditions and thus the method will give more than one answer. However, we will show that if $f$ exists, it is unique.
\begin{figure}
\includegraphics{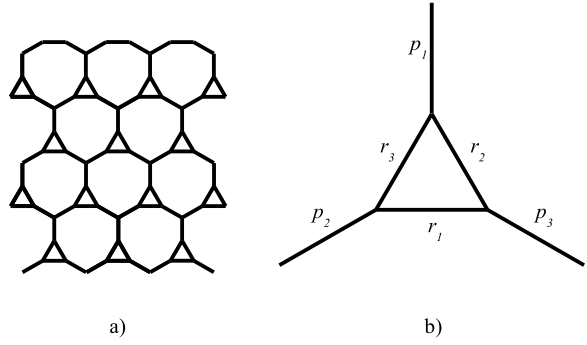}
\caption{a) The martini lattice; b) The assignment of probabilities on the unit cell.} \label{fig:martini}
\end{figure}
\begin{center}
\begin{figure}
 \includegraphics{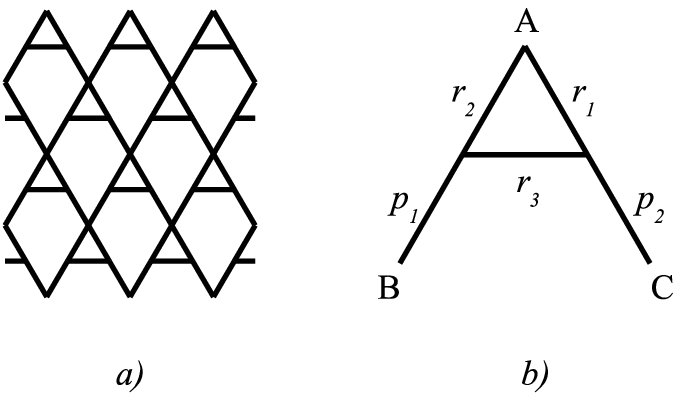}
\caption{a) The martini-A lattice; b) The assignment of probabilities on the unit cell.} \label{fig:Alattice}
\end{figure}
\end{center}
\begin{center}
\begin{figure}
 \includegraphics{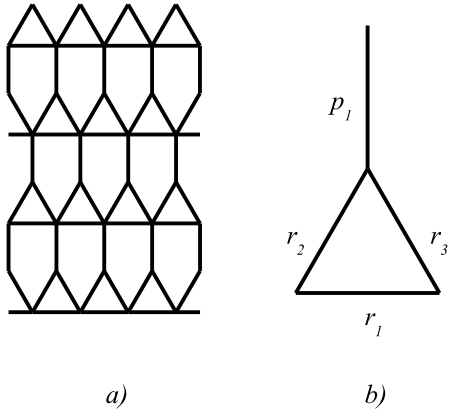}
\caption{a) The martini-B lattice; b) The assignment of probabilities on the unit cell.} \label{fig:Blattice}
\end{figure}
\end{center}
\section{Exact Solutions}
In what follows, we will need several exact solutions in order to have cases to which our approximations must reduce. We list a few of the important ones here.

\subsection{Martini lattice}
\begin{center}
\begin{figure}
 \includegraphics{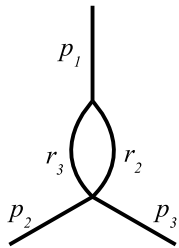} 
\caption{The unit cell resulting from setting $r_1=1$ in the martini lattice. This is just a honeycomb lattice with a complicated vertical bond.} \label{fig:comphex}
\end{figure}
\end{center}

We start with the martini lattice \cite{Fendley, Scullard06} (Figure \ref{fig:martini}a), and recover the previously reported inhomogeneous bond threshold \cite{ScullardZiff06,Wu06}. To solve this problem, we could use (\ref{eq:critcond}) with Figure \ref{fig:martini}b, which requires a rather 
involved enumeration of paths on the unit cell. However, by virtue of (\ref{eq:critcond}), and the fact that the probabilites $P$ are all linear functions of the bond probabilities, the critical function, denoted $\mathrm{M}(p_1,p_2,p_3,r_1,r_2,r_3)$, is also linear in all its arguments. Furthermore, we know that setting $r_1=0$ results in a honeycomb lattice in which two of the bonds are doubled in series, i.e.,
\begin{equation}
 \mathrm{M}(p_1,p_2,p_3,0,r_2,r_3)=\mathrm{H}(p_1,p_2 r_3, p_3 r_2) . \nonumber
\end{equation}
In order to satisfy this condition and linearity, the function must be of the form
\begin{equation}
\mathrm{M}(p_1,p_2,p_3,r_1,r_2,r_3)=\mathrm{H}(p_1,p_2 r_3, p_3 r_2)+r_1 \phi(p_1,p_2,p_3,r_2,r_3) \label{eq:Mwithphi}
\end{equation}
where $\phi$ is as-yet undetermined. Setting $r_1=1$ gives another honeycomb lattice, this time with a more complicated vertical bond (Fig \ref{fig:comphex}). The probability of crossing this bond is $p_1 [1-(1-r_2)(1-r_3)]$, i.e., the $p_1$ bond must be open and the $r_2$ and $r_3$ bonds cannot both be closed. Thus,
\begin{equation}
 \mathrm{M}(p_1,p_2,p_3,1,r_2,r_3)=\mathrm{H}(p_1 [1-(1-r_2)(1-r_3)],p_2,p_3) . \label{eq:Mcomplhex}
\end{equation}
Using (\ref{eq:Mcomplhex}) in (\ref{eq:Mwithphi}) allows us to determine $\phi$ uniquely. The resulting expression for $\mathrm{M}$ is
\begin{equation}
\mathrm{M}(p_1,p_2,p_3,r_1,r_2,r_3)=(1-r_1)\mathrm{H}(p_1,p_2 r_3, p_3 r_2)+r_1\mathrm{H}(p_1 [1-(1-r_2)(1-r_3)],p_2,p_3) \nonumber
\end{equation}
and expanding this out leads to the final result \cite{Wu06,ScullardZiff06},
\begin{eqnarray}
\mathrm{M}(p_1,p_2,p_3,r_1,r_2,r_3) &=&1 - p_1 p_2 r_3 - p_2 p_3 r_1 - p_1 p_3 r_2 - p_1 p_2 r_1 r_2 \nonumber \\
&-& p_1 p_3 r_1 r_3 - p_2 p_3 r_2 r_3 + p_1 p_2 p_3 r_1 r_2 \nonumber \\
&+& p_1 p_2 p_3 r_1 r_3 + p_1 p_2 p_3 r_2 r_3 + p_1 p_2 r_1 r_2 r_3 \nonumber \\
&+& p_1 p_3 r_1 r_2 r_3 + p_2 p_3 r_1 r_2 r_3 - 2 p_1 p_2 p_3 r_1 r_2 r_3=0 \ .
\label{eq:martini}
\end{eqnarray}
Solving the polynomial $\mathrm{M}(p,p,p,p,p,p)=0$ gives the exact homogeneous
bond percolation threshold, $p_c=1/\sqrt{2}$. Two special cases will also be useful later; the martini-A lattice (Figure \ref{fig:Alattice}) found by contracting the $p_3$ bond, $\mathrm{MA}(p_1,p_2,r_1,r_2,r_3)\equiv \mathrm{M}(p_1,p_2,1,r_1,r_2,r_3)$, and the martini-B lattice (Figure \ref{fig:Blattice}) which results from contraction of the $p_2$ and $p_3$ bonds, $\mathrm{MB}(p_1,r_1,r_2,r_3)\equiv \mathrm{M}(p_1,1,1,r_1,r_2,r_3)$.

The method we used to find this solution is somewhat less work than using (\ref{eq:critcond}), however its greatest virtue lies in the fact that it can be generalized to lattices that are not handled by (\ref{eq:critcond}). As we will show, the thresholds predicted in those cases are not exact but are always very good approximations, usually holding to $4$ or more decimal places.
\subsection{Stack of three triangles}
\begin{center}
\begin{figure}
 \includegraphics{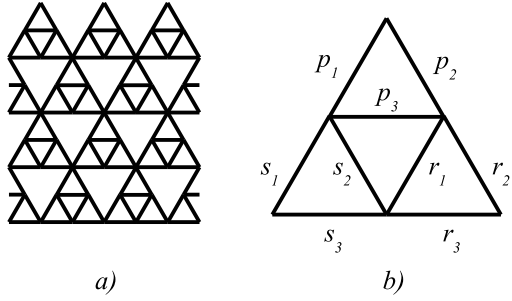} 
\caption{a) The stack of three triangles; b) the assignment of probabilities.} \label{fig:stackoftriangles}
\end{figure}
\end{center}

The stack of three triangles (Figure \ref{fig:stackoftriangles}) will be useful in deriving the approximation for the $(3,4,6,4)$ lattice. The unit cell has $9$ probabilities, and we denote its critical function by $\mathrm{SoT}(p_1,p_2,p_3,r_1,r_2,r_3,s_1,s_2,s_3)$ . This is a very long formula, and it is included as a text file among the source files of this submission. Here, we just report the polynomial for the homogeneous
case:
\begin{equation}
 \mathrm{SoT}(p,p,p,p,p,p,p,p,p)=1 - 3 p^2 - 9 p^3 + 3 p^4 + 45 p^5 - 72 p^6 + 45 p^7 - 12 p^8 + p^9=0\ . \nonumber
\end{equation}
The critical point of this lattice is $p_c=0.471629...$ \cite{Haji-Akbari}.
\subsection{Rocket lattice}
\begin{center}
\begin{figure}
 \includegraphics{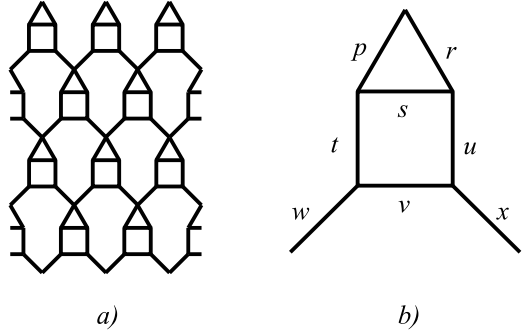} 
\caption{a) The ``rocket'' lattice; b) the assignment of probabilities.} \label{fig:rocket}
\end{figure}
\end{center}

This is another lattice that is helpful for the $(3,4,6,4)$ approximation. It is shown in Figure \ref{fig:rocket}, which we call ``rocket'' from the shape of the unit cell. The critical function is denoted $\mathrm{R}(p,r,s,t,u,v,w,x)$, and once again we will forego reporting the full threshold but have included it in the supplemental material. The result for the
homogeneous case is,
\begin{equation}
 1 - 3 p^3 - 4 p^4 + p^5 + 13 p^6 - 12 p^7 + 3 p^8=0\ . \nonumber
\end{equation}
The critical probability is $p_c=0.669182...$ .
\subsection{Decorated square}
\begin{center}
\begin{figure}
 \includegraphics{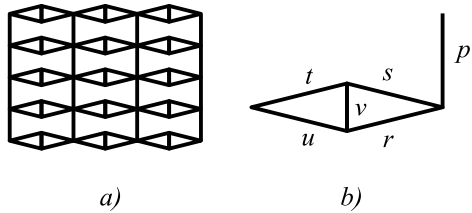} 
\caption{a) The ``decorated'' square lattice; b) the assignment of probabilities.} \label{fig:compsquare}
\end{figure}
\end{center}

This lattice (Figure \ref{fig:compsquare}) will be useful when we consider the $(3^3,4,3,4)$ lattice. It is really just a square lattice with a complicated horizontal bond. Its threshold is straightforward to find:
\begin{eqnarray}
 \mathrm{DS}(p, r, s, t, u, v)&\equiv& 1 - p - s t - r u  - r t v - s u v + r s t v \nonumber \\
&+& r s t u + r s u v + r t u v + s t u v - 2 r s t u v =0. \label{eq:decsquare}
\end{eqnarray}
\begin{center}
\begin{figure}
 \includegraphics{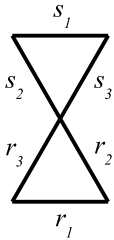} 
\caption{The assignment of probabilities on the kagome lattice unit cell.} \label{fig:kagome}
\end{figure}
\end{center}
\section{Approximations to unsolved problems}
We turn now to lattices whose thresholds are not known exactly. To derive these results we use the same reasoning as above, but because none of these cases is handled by (\ref{eq:critcond}), we would not expect any of them to be exact. This expectation is generally confirmed by comparison with precise numerical results. However, as we will show, the approximations they provide are consistently very good, and two interesting cases may turn out to be exact.

\subsection{Kagome lattice}
Bond percolation on the kagome lattice (Figure \ref{fig:archimedean}d) has been the subject of particular curiosity and numerous conjectures over the years \cite{Wu79,Tsallis,ZiffScullard06}. This is the one unknown threshold for the four most basic lattices that are generally studied (the square, triangular, honeycomb, and kagome). The attraction of this problem also stems from its similarity to the double-bond honeycomb lattice, which has an exactly known bond threshold \cite{SykesEssam}. Most of the attempted solutions have been based on finding a simple connection between the two, but none have been successful (although some have been quite close).

A shortcut like the one we used for the martini lattice is possible here. We assign the probabilities as shown in Figure \ref{fig:kagome}. If we set the $s_1$ bond probability to $0$ the result should be the martini-A lattice, so we must have
\begin{equation}
 \mathrm{K}(r_1,r_2,r_3,s_1,s_2,s_3)=\mathrm{MA}(s_3,s_2,r_2,r_3,r_1)+s_1 \phi(r_1,r_2,r_3,s_2,s_3) .\label{eq:kagomestep1}
\end{equation}
Setting this bond to $1$ gives the martini-B lattice with its $p_1$ bond doubled in parallel. We should therefore set $p_1=1-(1-s_2)(1-s_3)$ in the B lattice threshold:
\begin{equation}
 \mathrm{K}(r_1,r_2,r_3,1,s_2,s_3)=\mathrm{MB}(1-(1-s_2)(1-s_3),r_1,r_2,r_3) . \label{eq:kagomestep2}
\end{equation}
Setting $s_1=1$ in (\ref{eq:kagomestep1}) and substituting (\ref{eq:kagomestep2}) allows determination of $\phi(r_1,r_2,r_3,s_2,s_3)$, giving
\begin{equation}
 \mathrm{K}(r_1,r_2,r_3,s_1,s_2,s_3)=(1-s_1) \mathrm{MA}(s_3,s_2,r_2,r_3,r_1)+ s_1 \mathrm{MB}(1-(1-s_2)(1-s_3),r_1,r_2,r_3) . \label{eq:kagomeexp}
\end{equation}
After expanding, the result is:
\begin{eqnarray}
\mathrm{K}(r_1,r_2,r_3,s_1,s_2,s_3) &\equiv& 1 - r_1 s_1 - r_2 s_2 - r_3 s_3 - s_1 r_2 r_3- s_2 r_1 r_3 - s_3 r_1 r_2 \nonumber \\
&-& r_1 s_2 s_3 - r_2 s_1 s_3 - r_3 s_1 s_2 + s_1 r_1 r_2 r_3 + s_2 r_1 r_2 r_3 \nonumber \\
&+& s_3 r_1 r_2 r_3 + r_1 r_2 s_1 s_3 + r_1 r_3 s_1 s_2 + r_2 r_3 s_1 s_2  \nonumber \\
&+& r_2 r_3 s_1 s_3 + r_1 r_2 s_2 s_3 + r_2 s_1 s_2 s_3 + r_3 s_1 s_2 s_3 \nonumber \\
&+& r_1 r_3 s_2 s_3 + r_1 s_1 s_2 s_3 - r_1 r_2 r_3 s_1 s_3 - r_1 r_2 r_3 s_2 s_3 \nonumber \\
&-& r_1 r_2 r_3 s_1 s_2 - r_1 r_2 s_1 s_2 s_3 - r_1 r_3 s_1 s_2 s_3 \nonumber \\
&-& r_2 r_3 s_1 s_2 s_3 + r_1 r_2 r_3 s_1 s_2 s_3\ . \label{eq:kagome}
\end{eqnarray}
This inhomogeneous threshold was reported previously in
 \cite{ScullardZiff06}, but was derived by a different method. The critical probability is the solution to $\mathrm{K}(p,p,p,p,p,p)=0$, i.e.,
\begin{equation}
 1 - 3 p^2 - 6 p^3 + 12 p^4 - 6 p^5 + p^6=0
\end{equation}
with solution in $[0,1]$, $p_c=0.524430...$, which is identical to the
value conjectured by Wu in 1979 \cite{Wu79}. Two recent numerical estimates are $p_c = 0.52440499(2)$ \cite{Feng} and $p_c=0.52440510(5)$ \cite{Ding}, so we are close, but not quite right. The inhomogeneous threshold (\ref{eq:kagome}) is remarkable in that, absent a high-precision numerical result (which has only been available for the last twelve years \cite{ZiffSuding97}), it would be easy to convince oneself that it is the correct answer. It has all the necessary symmetry properties, and makes no incorrect prediction for any special case one can consider. Moreover, as is obvious from the derivation, it is the only function linear in its arguments that could possibly serve as the kagome threshold. The fact that it is incorrect means that the linearity assumption is wrong, but it is interesting that we can find such a function regardless, and that it makes a prediction so close to the correct answer. These properties are characteristic of all the criticality functions we will find here.

Equation (\ref{eq:kagomeexp}) has the formal appearance of an average of two lattice thresholds. This is the general form of a linear critical surface. If we have a lattice $L_1$ with critical surface $f_{L_1}(p_1,...,p_n)=0$, that reduces to $L_2$, which is a lattice with $m<n$ bonds in its unit cell and threshold $f_{L_2}(r_1,...,r_m)$, when $p_1=1$ and $L_3$, which has $k<n$ bonds in its unit cell and threshold $f_{L_3}(s_1,...,s_k)$, when $p_1=0$, then we have
\begin{eqnarray}
f_{L_1}(p_1,...,p_n)&=&p_1 f_{L_2}(\rho_1(p_2,...,p_n),...,\rho_m(p_2,...,p_n)) \nonumber \\
&+&(1-p_1) f_{L_3}(\eta_1(p_2,...,p_n),...,\eta_k(p_2,...,p_n)) \label{eq:generalexp}
\end{eqnarray}
where $\rho_i$ and $\eta_i$ are functions that are at most first-order in their arguments. $L_1$ is a kind of ``average'' of $L_2$ and $L_3$. Equation (\ref{eq:generalexp}) is the only way one can form a threshold that is linear in probabilities and that correctly reduces to $L_2$ and $L_3$ in the appropriate limits. This indicates that, in general, if a linear function exists that makes correct predictions in all special cases, then it is the {\it only} such function and the method of obtaining it is irrelevant.
\begin{center}
\begin{figure}
 \includegraphics{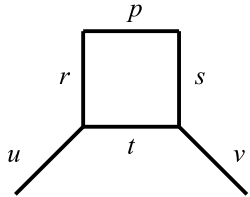} 
\caption{Assignment of probabilities on the $(4,8^2)$ lattice.} \label{fig:foureightsquared}
\end{figure}
\end{center}
\subsection{$(4,8^2)$ lattice}
The basic cell of the $(4,8^2)$ (``four-eight'' or ``FE'' lattice) lattice has six bonds and is shown in Figure \ref{fig:foureightsquared} along with the probability assignments. Removal of the $t$ bond results in the honeycomb lattice with two doubled bonds, and setting $t=1$ gives the martini-A lattice. Equation (\ref{eq:generalexp}) implies
\begin{equation}
 \mathrm{FE}(p,r,s,t,u,v)=(1-t) \mathrm{H}(p,ru,sv)+t \mathrm{MA}(u,v,r,s,p) .\nonumber
\end{equation}
Expanding gives the complete function, which will also be useful later,
\begin{eqnarray}
\mathrm{FE}(p,r,s,t,u,v)&=&1 - (p r u + s t u + p s v + r t v) - (r s u v + p t u v) +\nonumber \\
p r s u v + p t r u v &+& p s t u v + r s t u v + p r s t u + p r s t v - 2 p r s t u v =0 \nonumber .\label{eq:foureightsquared}
\end{eqnarray}
The critical polynomial $\mathrm{FE}(p,p,p,p,p,p)=0$ is then
\begin{equation}
 1 - 4 p^3 - 2 p^4 + 6 p^5 - 2 p^6=0 \nonumber
\end{equation}
with solution on $[0,1]$, $p_c=0.676835...$ . According to Parviainen \cite{Parviainen}, this lattice has $p_c=0.676802...$ with standard error $6.3 \times 10^{-7}$. Although our answer lies outside the error bars, it does agree to four significant figures. It also lies well within the confidence interval of Riordan and Walters \cite{Riordan} shown in Table \ref{table:bondthresholds}.

\begin{center}
\begin{figure}
 \includegraphics{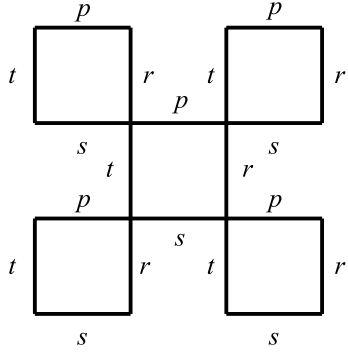} 
\caption{The square lattice with the checkerboard assignment of probabilities.} \label{fig:checkerboard}
\end{figure}
\end{center}

This threshold makes another interesting prediction. Setting $u=v=1$ gives the square lattice with four different probabilities as shown in Figure \ref{fig:checkerboard}, and we have
\begin{equation}
\mathrm{CB}(p,r,s,t)\equiv 1 - p r - p s - r s - p t - r t - s t + p r s + p r t + p s t + r s t =0 . \label{eq:checkerboard}
\end{equation}
This arrangement was named the ``checkerboard'' by Wu \cite{Wu82} and he conjectured its threshold for the $q-$state Potts model based on symmetry and special cases, in much the same way as we are doing here. Our result is identical with the $q \rightarrow 1$ limit of Wu's formula. Although his conjecture was shown to be incorrect for $q=3$ \cite{Enting}, our simulations suggest that it may be exact for percolation. However, proving this will require a new technique, as it is not handled by (\ref{eq:critcond}).
\begin{center}
\begin{figure}
 \includegraphics{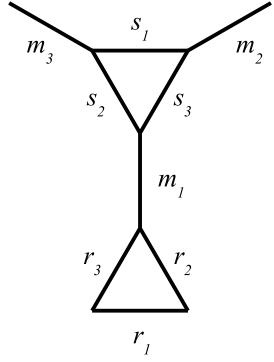} 
\caption{The assignment of probabilities on the $(3,12^2)$ unit cell.} \label{fig:threetwelvesquared}
\end{figure}
\end{center}
\subsection{$(3,12^2$) lattice}
The approximation for the $(3,12^2)$ lattice (often called the 3-12 lattice \cite{MaillardRolletWu}) was also previously mentioned by the authors \cite{ScullardZiff06}. We will just repeat the result here, which we write in a factorized form to save space,
\begin{eqnarray}
1&-& m_1 m_2 ( r_3 + r_1 r_2 - r_1 r_2 r_3) (s_3 + s_1 s_2 - s_1 s_2 s_3) \nonumber \\
&-& m_1 m_3 (r_2 + r_1 r_3 - r_1 r_2 r_3) (s_2 + s_1 s_3 - s_1 s_2 s_3) \nonumber \\
&-& m_2 m_3 (r_1 + r_2 r_3 - r_1 r_2 r_3) (s_1 + s_2 s_3 - s_1 s_2 s_3) \nonumber \\
&+& m_1 m_2 m_3 (r_1 r_2 + r_1 r_3 + r_2 r_3 - 2 r_1 r_2 r_3) \nonumber \\
&\times&(s_1 s_2 + s_1 s_3 + s_2 s_3 -
2 s_1 s_2 s_3)=0\ , \label{eq:3_12}
\end{eqnarray}
where we have used the probability assignments from Figure \ref{fig:threetwelvesquared}. The polynomial for the homogeneous case factorizes to
\begin{equation}
(1 + p - 2 p^3 + p^4)(1 - p + p^2 + p^3 - 7 p^4 + 4 p^5)=0 \ . \nonumber
\end{equation}
The part that contains the solution in $[0,1]$ is
\begin{equation}
1 - p + p^2 + p^3 - 7 p^4 + 4 p^5=0
\end{equation}
and yields $p_c=0.7404233179...$ . According to Ding et al. \cite{Ding},
$p_c(3,12^2)=0.7404207(2)$, so our approximation agrees to $5$ significant figures.
\begin{center}
\begin{figure}
 \includegraphics{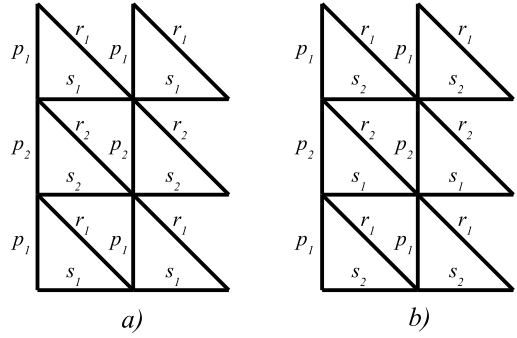} 
\caption{a) The ``striped'' triangular lattice; b) ``re-partioned'' version of a). The critical threshold should be invariant under this permutation of the probabilities.} \label{fig:sttr}
\end{figure}
\end{center}
\subsection{$(3^3,4^2)$ lattice}
\begin{center}
\begin{figure}
 \includegraphics{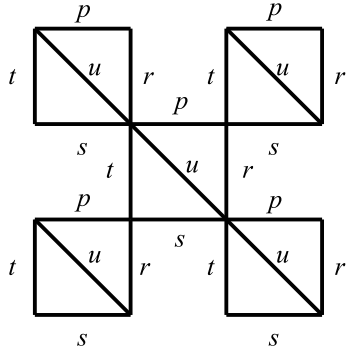} 
\caption{The 5-bond bow-tie lattice. The predicted threshold for this situation, equation (\ref{eq:bowtie}), is possibly exact.} \label{fig:bowtie5}
\end{figure}
\end{center}

For the remaining lattices, we must employ a more brute-force method to find the critical function, requiring the use of Mathematica to perform the algebra. For the $(3^3,4^2)$ lattice, we consider an inhomogeneous triangular lattice, which we call the ``striped'' triangular lattice because the probabilities are assigned in distinct stripes (Figure \ref{fig:sttr}a). The $(3^3,4^2)$ lattice appears when $r_1=0$. We start by defining the most general function that is at most first-order in its variables,
\begin{equation}
 f(p_1,r_1,s_1,p_2,r_2,s_2)=\sum_{i=0}^1 ... \sum_{n=0}^1 a_{ijklmn} p_1^i r_1^j s_1^k p_2^l r_2^m s_2^n
\end{equation}
where there are $64$ $a$'s to be determined (in general, there will be $2^n$ coefficients, where $n$ is the number of bonds in a unit cell). However, we are free to set the constant term to $1$, because we can multiply $f=0$ by any constant and not change the threshold. Now we impose all the known symmetries and special cases onto $f$. For example, simply re-partitioning the lattice as shown in Figure \ref{fig:sttr}b should not change the result, so
\begin{equation}
 f(p_1,r_1,s_1,p_2,r_2,s_2)=f(p_1,r_1,s_2,p_2,r_2,s_1) . \nonumber
\end{equation}
This constraint tells us that a certain set of the $a$'s are equal to each other, reducing the number of coefficients by 16 in this case. Similarly, the threshold should be invariant under a flip about a vertical axis:
\begin{equation}
 f(p_1,r_1,s_1,p_2,r_2,s_2)=f(r_1,p_1,s_1,r_2,p_2,s_2) \nonumber
\end{equation}
which eliminates another 18 coefficients. Next, there is the symmetry between the $1$ and $2$ bonds,
\begin{equation}
f(p_1,r_1,s_1,p_2,r_2,s_2)=f(p_2,r_2,s_2,p_1,r_1,s_1) \nonumber
\end{equation}
eliminating $9$ $a$'s, leaving $20$ to be determined. Setting $r_1=s_1=0$ should give the threshold for the martini-B lattice,
\begin{equation}
 f(p_1,0,0,p_2,r_2,s_2)=\mathrm{MB}(p_1,s_2,p_2,r_2) \nonumber
\end{equation}
leaving $9$ undetermined coefficients. Because we are dealing with the triangular lattice $p_1=p_2=p, r_1=r_2=r, s_1=s_2=s$ should yield (\ref{eq:triangular}). The function $f$ will not simply reduce to the function $\mathrm{T}(p,r,s)$ because $f(p,r,s,p,r,s)$ has second-order terms. Rather, $\mathrm{T}(p,r,s)$ must factor out of the resulting expression. Because $\mathrm{T}(p,r,s)=0$ is equivalent to $p=(r+s-1)/(rs-1)$ we must have
\begin{equation}
f \left(\frac{r+s-1}{rs-1},r,s,\frac{r+s-1}{rs-1},r,s \right)=0 . \nonumber
\end{equation}
This leaves only one coefficient unconstrained. We find this by observing that $r_1=r_2=0$ gives the same lattice as setting $r_1=p_2=0$, with some probabilities changed, i.e.,
\begin{equation}
 f(p_1,0,s_1,p_2,0,s_2)=f(p_1,0,s_1,0,p_2,s_2) .\nonumber
\end{equation}
This finally fixes the threshold, which is given by
\begin{eqnarray}
\mathrm{ST}(p_1,r_1,s_1,p_2,r_2,s_2)&\equiv&1 - s_1- s_2 - p_1 p_2- r_1 r_2 - p_1 r_1 - p_2 r_1 - p_2 r_2 - p_1 r_2 \nonumber \\
 &+& s_1 s_2 + p_1 p_2 r_2 + p_1 p_2 r_1 +p_1 r_1 r_2 + p_2 r_1 r_2 + p_1 r_1 s_1 \nonumber \\
 &+& p_2 r_2 s_1 +p_1 r_1 s_2 + p_2 r_2 s_2 - p_1 r_1 s_1 s_2 - p_2 r_2 s_1 s_2   \nonumber \\
 & -& p_1 p_2 r_1 r_2 s_2- p_1 p_2 r_1 r_2 s_1 + p_1 p_2 r_1 r_2 s_1 s_2=0 \label{eq:sttr}
\end{eqnarray}
The $(3^3,4^2)$ lattice is obtained by setting $r_1=0$:
\begin{eqnarray}
\mathrm{TF}(p_1,p_2,r_2,s_1,s_2)&\equiv& 1 - s_1- s_2 + s_1 s_2 - p_1 p_2 - p_1 r_2 - p_2 r_2   \nonumber \\ 
&+& p_2 r_2 s_1 + p_2 r_2 s_2 + p_1 p_2 r_2 - p_2 r_2 s_1 s_2 = 0\label{eq:tcfs}
\end{eqnarray}
The homogeneous polynomial is
\begin{equation}
 \mathrm{TF}(p,p,p,p,p)=1 - 2 p - 2 p^2 + 3 p^3 - p^4=0 \nonumber
\end{equation}
with solution on $[0,1]$ $p_c=0.419308...$ . Parviainen finds $p_c \approx 0.41964191$ with standard error $4.3 \times 10^{-7}$, a difference of $0.00033$. Once again, our result is close but not in agreement with the numerical value. It may seem somewhat surprising that we could satisfy all our various criteria using only a function first-order in probabilities. What is more, we have a formula that makes predictions for cases that we did not explicitly consider in the derivation. For example, setting $s_1=0$ gives an inhomogeneous version of the bow-tie lattice \cite{Wierman84}, $\mathrm{BT}(p,r,s,t,u)\equiv \mathrm{ST}(t,s,0,r,p,u)$ (Figure \ref{fig:bowtie5}), so
\begin{eqnarray}
 \mathrm{BT}(p,r,s,t,u) &=& 1 - u - p r - p s - r s - p t \nonumber \\
 &-& r t - s t + p r s + p r t  + p s t \nonumber \\
 &+& r s t+ p r u + s t u - p r s t u = 0 .\label{eq:bowtie}
\end{eqnarray}
Setting the probabilities equal gives
\begin{equation}
 1 - p - 6 p^2 + 6 p^3 - p^5=0
\end{equation}
with $p_c^{\mathrm{bowtie}}=0.404518...$, which is the exact answer found by Wierman \cite{Wierman84}. Based upon numerical studies, we believe that the more general equation (\ref{eq:bowtie}) may also be exact \cite{ZiffScullard10}, even though it is not obvious how to derive it using the normal duality arguments.

\begin{center}
\begin{figure}
 \includegraphics{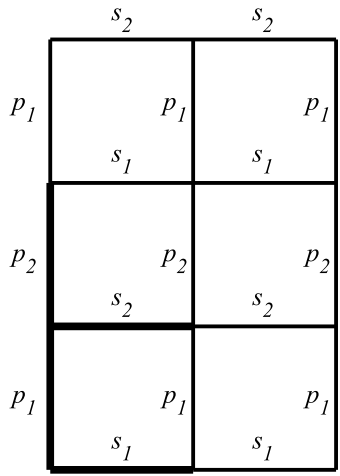} 
\caption{The lattice resulting from setting $r_1=r_2=0$ in (\ref{eq:sttr}) with the unit cell highlighted. Equation (\ref{eq:secsquare}) is the predicted threshold.} \label{fig:secsquare}
\end{figure}
\end{center}

Another interesting prediction is found by setting $r_1=r_2=0$ (Fig \ref{fig:secsquare}). This again results in a square lattice, but the probabilities are assigned differently from the checkerboard. We call this the ``striped'' square, and the threshold is predicted by (\ref{eq:sttr}) to be
\begin{equation}
 -p_1 p_2+s_1 s_2 -s_1-s_2+1=0 . \label{eq:secsquare}
\end{equation}
This formula has several interesting properties. For example, if we set $p_1=p_2=p$ and $s_1=s_2=s$ we have the standard square lattice and (\ref{eq:secsquare}) can be factored to give
\begin{equation}
 (1 + p - s) (1 - p - s)=0 \nonumber
\end{equation}
leading to the correct threshold $1 - p - s=\mathrm{S}(p,s)=0$. We can discard the first term in brackets because $s=1+p$ obviously does not give answers in $(0,1)$. Although the underlying lattice is self-dual, this assignment of probabilities is not. The dual lattice is again a square, but the inequivalent $p$ and $s$ bonds are interchanged. Our formula reflects this property too, because if we set $p_i \rightarrow 1-p_i$ and $s_i \rightarrow 1-s_i$ we indeed end up with
\begin{equation}
-s_1 s_2+p_1 p_2 -p_1-p_2+1=0 .\nonumber
\end{equation}
Finally, setting all probabilities equal leads to $p_c=1/2$ as required for the square lattice. Although this formula rather nicely exhibits all the required features, it does not turn out to be exact. This is easily seen by a numerical example. If we set $s_1=s_2=p_2=0.4$, then (\ref{eq:secsquare}) predicts $p_1=0.9$. However, our numerical work \cite{ZiffScullard10} gives a value of $p_1 \approx 0.9013$, which rules out this prediction.

\begin{center}
\begin{figure}
 \includegraphics{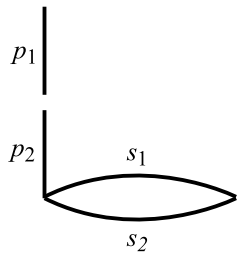} 
\caption{Unit cell of the square lattice for which (\ref{eq:secsquare}) is the exact threshold.} \label{fig:secsquareexact}
\end{figure}
\end{center}

Finally, we point out that the path we took to constrain $\mathrm{ST}(p_1,r_1,s_1,p_2,r_2,s_2)$ might not be the most efficient or direct; the order in which we apply the conditions is mostly just the order in which we thought of them. However, as we pointed out above, once one has found a consistent linear threshold it is the only function that has all the correct properties. The converse is not true however --- two different lattices may be self-consistently represented by the same linear function. An example is equation (\ref{eq:secsquare}), which is also the exact threshold of the square lattice with vertical bonds replaced by two in series, and horizontal bonds replaced by two in parallel (Figure (\ref{fig:secsquareexact})), i.e. $\mathrm{S}(p_1 p_2,1-(1-s_1)(1-s_2))$.
\begin{center}
\begin{figure}
 \includegraphics{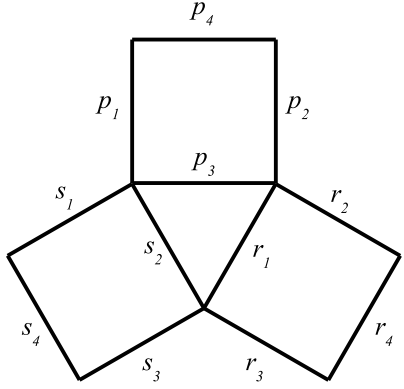} 
\caption{Probabilities for the $(3,4,6,4)$ lattice.} \label{fig:three464}
\end{figure}
\end{center}
\subsection{$(3,4,6,4)$ lattice}
The basic cell for the $(3,4,6,4)$ lattice of Figure \ref{fig:archimedean}h is shown in Figure \ref{fig:three464}. It has $12$ bonds, making it the largest cell we will consider. There are $4095$ coefficients that must be constrained (the constant is set to $1$ as usual). The constraints that fully determine the critical function are as follows:
\begin{enumerate}
\item We can rotate the unit cell by $120$ degrees without changing the threshold, meaning that we have
\begin{equation}
f(p_1,p_2,p_3,p_4,r_1,r_2,r_3,r_4,s_1,s_2,s_3,s_4)=f(r_2,r_3,r_1,r_4,s_2,s_3,s_1,s_4,p_2,p_3,p_1,p_4) . \nonumber
\end{equation}
Rotating the opposite direction does not give any additional constraint.
\item It is possible to re-partition the lattice, as we did in Figure \ref{fig:sttr} for the striped triangular situation, so the unit cell is still a set of three squares but the probabilities are shuffled, i.e.,
\begin{equation}
 f(p_1, p_2, p_3, p_4, r_1, r_2, r_3, r_4, s_1, s_2, s_3, s_4) = 
 f(s_1, s_3, s_4, s_2, p_4, p_2, p_1, p_3, r_2, r_4, r_3, r_1) .\nonumber
\end{equation}
\item We can reflect the lattice about a vertical line drawn through the middle of the unit cell, so
\begin{equation}
 f(p_1, p_2, p_3, p_4, r_1, r_2, r_3, r_4, s_1, s_2, s_3, s_4) = 
 f(p_2, p_1, p_3, p_4, s_2, s_1, s_3, s_4, r_2, r_1, r_3, r_4) .\nonumber
\end{equation}
Reflections about other axes do not give any further constraints.
\item Contracting the $p_4,r_4,s_4$ bonds, i.e., setting their probabilities to $1$, results in the stack of triangles, so
\begin{equation}
 f(p_1, p_2, p_3, 1, r_1, r_2, r_3, 1, s_1, s_2, s_3, 1) = 
 \mathrm{SoT}(p_1, p_2, p_3, r_1, r_2, r_3, s_1, s_2, s_3) .\nonumber
\end{equation}
\item Setting $s_3,r_3=0$ results in the rocket lattice (Figure \ref{fig:rocket}) with a complicated $v$ bond:
\begin{equation}
 f(p_1, p_2, p_3, p_4, r_1, r_2, 0, r_4, s_1, s_2, 0, s_4) = 
 \mathrm{R}(r_4, s_4, p_4, p_1, p_2, 1 - (1 - p_3)(1 - s_2r_1), s_1, r_2)) \nonumber
\end{equation}
\item If we set $r_2=r_4=s_2=s_3=0$, the $(4,8^2)$ lattice with some doubled bonds results. Because we previously found the only linear function of probabilities that would qualify as that threshold, (\ref{eq:foureightsquared}), the current function must reduce to that one. That is,
\begin{equation}
 f(p_1, p_2, p_3, p_4, r_1, 0, r_3, 0, s_1, 0, 0, s_4) = 
 \mathrm{FE}(p_4, p_1, p_2, p_3, s_1 s_4, r_1 r_3) \nonumber
\end{equation}
\item A version of the kagome lattice results when we remove the $p_3,r_1$, and $s_2$ bonds. Once again, the only consistent linear threshold is given by (\ref{eq:kagome}) so we expect reduction to that formula:
\begin{equation}
 f(p_1, p_2, 0, p_4, 0, r_2, r_3, r_4, s_1, 0, s_3, s_4) = 
 \mathrm{K}(p_4, s_4, r_4, s_3 r_3, p_2 r_2, s_1 p_1) . \nonumber
\end{equation}
\item Setting $r_2=r_3=s_1=s_3=1$ gives a $(3^3,4^2)$ with several decorated bonds. This requires
\begin{eqnarray}
 &f(p_1, p_2, p_3, p_4, r_1, 1, 1, r_4, 1, s_2, 1, s_4) =& \nonumber \\ 
 &\mathrm{TF}(1 - (1 - p_1)(1 - p_2), p_3, 1 - (1 - r_1)(1 - r_4), 1 - (1 - s_2)(1 - s_4), p_4) .& \nonumber
\end{eqnarray}
\item Setting $p_1=p_2=r_3=s_1=1$ and $s_4=0$ gives a version of the dual to the $(4,8^2)$ lattice. As such we will need
\begin{equation}
 \mathrm{FED}(p,r,s,t,u,v)\equiv-\mathrm{FE}(1-p,1-r,1-s,1-t,1-u,1-v) . \nonumber
\end{equation}
and so
\begin{eqnarray}
 &f(1, 1, p_3, p_4, r_1, r_2, 1, r_4, 1, s_2, s_3, s_4)=& \nonumber \\
 &\mathrm{FED}(s_4, 1 - (1 - p_3)(1 - p_4), 1 - (1 - r_2)(1 - s_3), r_1, s_2, r_4)& \nonumber
\end{eqnarray}
\item We can get a version of the honeycomb lattice by assigning $p_1=r_1=r_2=s_3=s_4=0$:
\begin{equation}
 f(0, p_2, p_3, p_4, 0, 0, r_3, r_4, s_1, s_2, 0, 0) = 
 \mathrm{H}(p_2 p_3 p_4, s_1 r_4, s_2 r_3) \nonumber
\end{equation}
\item We now have only one coefficient left. Setting $p_3=p_4=0$ results in a 10-bond version of the dual to the bow-tie lattice, with the known threshold $p_c=0.595482...$ . Requiring
\begin{equation}
 f(p_c,p_c,0,0,p_c,p_c,p_c,p_c,p_c,p_c,p_c,p_c)=0
\end{equation}
fixes the final coefficient.
\end{enumerate}
All coefficients are now constrained, and the full expression, containing $879$ terms, is included in the supplemental material. The polynomial for this lattice is
\begin{equation}
 1 - 6 p^3 - 12 p^4 - 6 p^5 + 69 p^6 + 60 p^7 - 363 p^8 + 448 p^9 - 
 252 p^{10} + 66 p^{11} - 6 p^{12} =0 \nonumber
\end{equation}
with $p_c=0.524821111889...$ . Parviainen's numerical result is $p_c=0.52483258$ with standard error $5.3 \times 10^{-7}$. Once again we are outside the error bars but the difference is only $0.000012$ .

\begin{center}
\begin{figure}
 \includegraphics{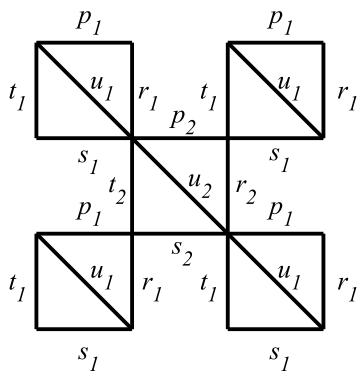} 
\caption{The bow-tie lattice with $10$ different probabilities.} \label{fig:bowtie10}
\end{figure}
\end{center}

\begin{center}
\begin{figure}
 \includegraphics{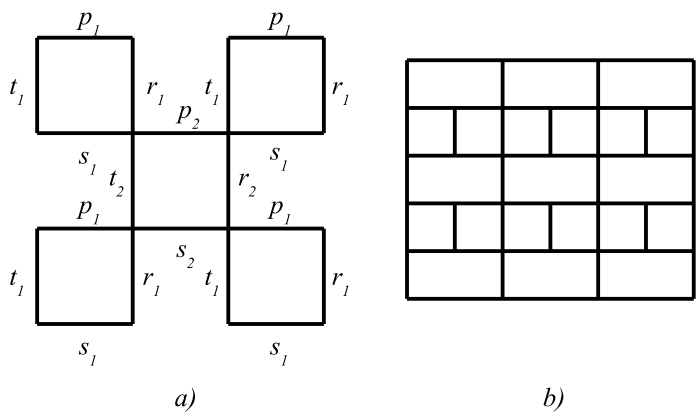} 
\caption{a) The extended checkerboard lattice with 8 independent probabilities; b) The result of setting $r_1=0$. We use this lattice to test our prediction for a).} \label{fig:ext_checkerboard}
\end{figure}
\end{center}

As with the $(3^3,4^2)$ lattice, we have a few other predictions of this formula that are worth discussing. As mentioned in step $9$, setting $p_3=p_4=0$ gives a 10-probability version of the bow-tie dual lattice, from which the 10-probability bow-tie (Fig \ref{fig:bowtie10}) threshold, $\mathrm{EBT}(p_1,r_1,s_1,t_1,u_1,p_2,r_2,s_2,t_2,u_2)$, can be found. As usual, all special cases seem to have the right properties. To test whether this inhomogeneous threshold is actually exact, we will reduce it to the extended checkerboard, i.e., the square lattice with $8$ different probabilities (Fig \ref{fig:ext_checkerboard}a), by setting $p_1=p_2=1$ and $p_3=p_4=0$. So 
\begin{equation}
\mathrm{ECB}(p_1,r_1,s_1,t_1,p_2,r_2,s_2,t_2)\equiv f(1,1,0,0,p_2,s_2,r_2,t_2,p_1,s_1,t_1,r_1) . \nonumber
\end{equation}
Now, setting $r_1=0$ and the other probabilities equal, we find a lattice with an unknown threshold (Fig \ref{fig:ext_checkerboard}b). The polynomial is
\begin{equation}
 1 - 3 p^2 - 5 p^4 + 12 p^5 - 7 p^6 + p^7 =0 \nonumber
\end{equation}
with solution on $[0,1]$ $p_c=0.5696764123...$ . Numerical simulation of this lattice gives $p_c \approx 0.56982$ \cite{ZiffScullard10}, so unfortunately the extended bow-tie and checkerboard thresholds predicted here cannot be exact.

\begin{center}
\begin{figure}
 \includegraphics{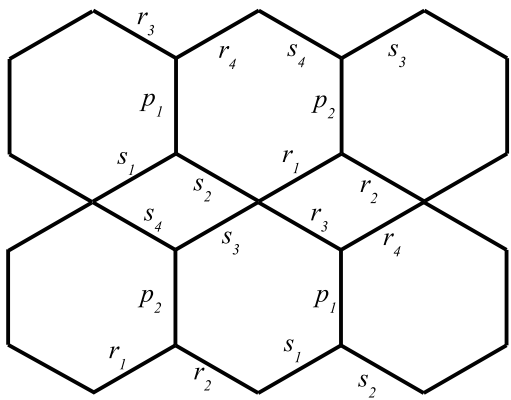} 
\caption{The lattice that results when setting $p_3=p_4=0$ in Figure \ref{fig:three464}. This is the dual of the bow-tie lattice with $10$ different probabilities.} \label{fig:bowtiedual10}
\end{figure}
\end{center}
\begin{center}
\begin{figure}
 \includegraphics{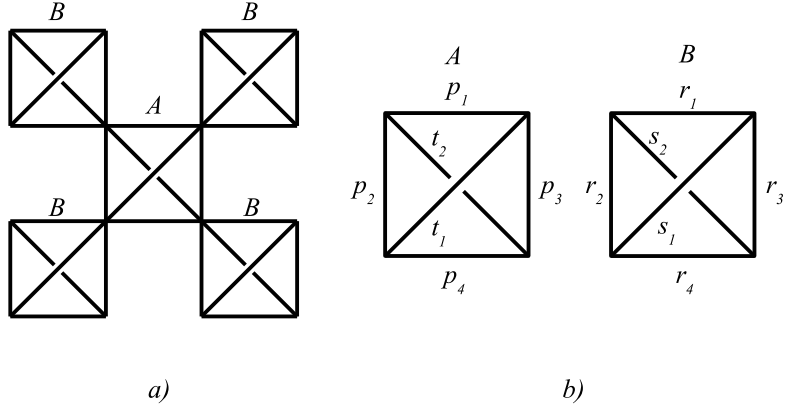} 
\caption{a) The line graph of the square lattice, which is used to derive the $(3^2,4,3,4)$ polynomial; b) The assignment of probabilities.} \label{fig:sqcovering}
\end{figure}
\end{center}
\subsection{$(3^2,4,3,4)$ lattice}
\begin{center}
\begin{figure}
 \includegraphics{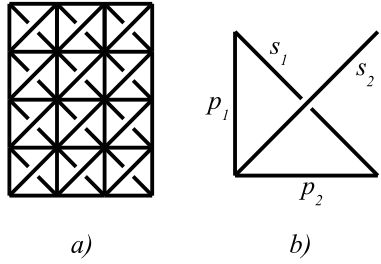} 
\caption{a) The square matching graph. The crossed bonds are not connected inside the square; b) The assignment of probabilities.} \label{fig:sqmatching}
\end{figure}
\end{center}

The final case we consider is the $(3^2,4,3,4)$ lattice. To do this, we start with the lattice shown in Figure \ref{fig:sqcovering} which is the covering graph of the square lattice with $12$ probabilities. We will constrain it until we have a sufficient number of coefficients for the lattice we want, which appears when $s_1=t_2=0$. An approximation that is helpful here is the one for the 4-bond square matching (SM) lattice (Figure \ref{fig:sqmatching}). By now it should be clear how this is done, so we only state the result:
\begin{eqnarray}
 \mathrm{SM}(p_1,p_2,s_1,s_2)&\equiv&1 - p_1 - p_2 - s_1 + p_1 p_2 s_1 - s_2 \nonumber \\
&+& p_1 p_2 s_2 + p_1 s_1 s_2 + p_2 s_1 s_2 - 2 p_1 p_2 s_1 s_2=0 \label{eq:sqmatch}
\end{eqnarray}
For uniform probabilities we have
\begin{equation}
\mathrm{SM}(p,p,p,p)= 1 - 4 p + 4 p^3 - 2 p^4=0 \nonumber
\end{equation}
with solution $p_c=0.266385...$, which of course is not exact \cite{ZiffScullard10}.

We will now simply list the constraints used in deriving the approximation to the $(3^2,4,3,4)$ lattice.
\begin{enumerate}
 \item Rotation:
\begin{equation} f(p_1,p_2,p_3,p_4,r_1,r_2,r_3,r_4,s_1,s_2,t_1,t_2)=f(p_2,p_4,p_1,p_3,r_2,r_4,r_1,r_3,s_2,s_1,t_2,t_1) \nonumber
\end{equation}
\item Reflection:
\begin{equation}
 f(p_1,p_2,p_3,p_4,r_1,r_2,r_3,r_4,s_1,s_2,t_1,t_2)=f(r_2,r_1,r_4,r_3,p_2,p_1,p_4,p_3,t_1,t_2,s_1,s_2) \nonumber
\end{equation}
\item Re-partitioning:
\begin{equation}
 f(p_1,p_2,p_3,p_4,r_1,r_2,r_3,r_4,s_1,s_2,t_1,t_2)=f(r_1,r_2,r_3,r_4,p_1,p_2,p_3,p_4,t_1,t_2,s_1,s_2) \nonumber
\end{equation}
\item 
\begin{eqnarray}
 &f(p_1, p_2, p_3, p_4, r_1, r_2, r_3, r_4, s_1, 0, t_1, 0)& = \nonumber \\
 &\mathrm{EBT}(p_1, p_2, p_3, p_4, r_1, r_2, r_3, r_4, s_1, t_1)& \nonumber
\end{eqnarray}
\item 
\begin{eqnarray}
 &f(1, p_2, p_3, p_4, 1, r_2, r_3, r_4, s_1, s_2, t_1, t_2) = & \nonumber \\
 &\mathrm{ST}(p_*, r_*, r_4, s_*, t_*, p_4)& \nonumber
\end{eqnarray}
where $p_* = 1 - (1 - t_1)(1 - r_3)$, $r_* = 1 - (1 - r_2)(1 - t_2)$, $s_*=1 - (1 - s_1)(1 - p_3)$, and $t_* =  1 - (1 - p_2)(1 - s_2)$ .
\item Using (\ref{eq:decsquare}):
\begin{eqnarray}
 &f(p_1, p_2, p_3, 0, r_1, r_2, r_3, 1, 0, 0, t_1, t_2) =& \nonumber \\ 
 &\mathrm{DS}(p_1, p_2, p_3, r_1, 1 - (1 - t_1)(1 - r_2),
  1 - (1 - t_2)(1 - r_3))& \nonumber
\end{eqnarray}
\item Using (\ref{eq:tcfs}):
\begin{eqnarray}
 &f(p_1, p_2, p_3, p_4, r_1, 1, 1, r_4, s_1, 0, t_1, t_2) =& \nonumber \\
 & \mathrm{TF}(1 - (1 - r_1)(1 - t_1)(1 - t_2)(1 - r_4), p_3, 
   1 - (1 - p_1)(1 - p_4), s_1, p_2)& \nonumber
\end{eqnarray}
\item Now we need (\ref{eq:sqmatch}):
\begin{eqnarray}
 &f(p_1, p_2, p_3, p_4, 1, 1, 1, r_4, s_1, s_2, t_1, t_2) = & \nonumber \\
  &\mathrm{SM}(1 - (1 - p_2)(1 - p_3), 1 - (1 - p_1)(1 - p_4), s_1, 
   s_2))& \nonumber
\end{eqnarray}
\item 
\begin{eqnarray}
 &f(p_1, p_2, p_3, p_4, r_1, r_2, 1, r_4, s_1, s_2, 1, 1) =& \nonumber \\ 
 &\mathrm{SM}(1 - (1 - p_2)(1 - p_3), 1 - (1 - p_1)(1 - p_4), s_1, 
   s_2)& \nonumber
\end{eqnarray}
\item
\begin{eqnarray}
 &f(p_1, p_2, p_3, p_4, 1, r_2, r_3, 1, s_1, s_2, t_1, 1) =& \nonumber \\
 & \mathrm{SM}(1 - (1 - p_2)(1 - p_3), 1 - (1 - p_1)(1 - p_4), s_1, 
   s_2)& \nonumber
\end{eqnarray}
\item
\begin{eqnarray}
 &f(p_1, p_2, p_3, p_4, 1, r_2, 1, r_4, s_1, s_2, t_1, 1) =& \nonumber \\ 
 &\mathrm{SM}(1 - (1 - p_2)(1 - p_3), 1 - (1 - p_1)(1 - p_4), s_1, s_2)& \nonumber
\end{eqnarray}
\item
\begin{equation}
 f(p_1, p_2, p_3, p_4, r_1, r_2, r_3, r_4, 0, 1, 1, 0) = \mathrm{CB}(\bar{p}, \bar{r}, \bar{s},\bar{t}) \nonumber
\end{equation}
where $\bar{p}=1 - (1 - r_2)(1 - r_4)$, $\bar{r}=1 - (1 - p_3)(1 - p_4)$, $\bar{s}=1 - (1 - p_1)(1 - p_2)$, and $\bar{t}=1 - (1 - r_1)(1 - r_3)$.
\end{enumerate}
This finally constrains the coefficients we need to get the critical surface for the $(3^2,4,3,4)$ lattice, which has $458$ separate terms and can be found in the supplemental material. The homogeneous polynomial is
\begin{equation}
 1 - 4 p^2 - 12 p^3 - 2 p^4 + 106 p^5 - 186 p^6 + 132 p^7 - 36 p^8 - 
 2 p^9 + 2 p^{10}=0 \nonumber
\end{equation}
or $p_c=0.414120304...$ . Parviainen's numerical result is $p_c \approx 0.41413743$, with standard error $4.6 \times 10^{-7}$. Although outside the error bars, our prediction differs by only $0.000017$.
\begin{table}
\begin{center}
\begin{tabular}{clclclclc}
lattice & & $p_c^{\mathrm{approx}}$ & & $p_c^{\mathrm{num}}$& & bounds & & $|p_c^{\mathrm{approx}}-p_c^{\mathrm{num}}|$\\
\hline
kagome &\vline& $0.5244297$&\vline&$0.5244050$&\vline&$[0.52415,0.52465]$&\vline&$2.5 \times 10^{-5}$\\
$(3,12^2)$ &\vline& $0.7404233$ &\vline& $0.7404207$&\vline&$[0.7402,0.7407]$&\vline&$2.6 \times 10^{-6}$\\
$(4,8^2)$ &\vline& $0.676835$&\vline&$0.676802$&\vline&$[0.6766,0.6770]$&\vline&$3.3 \times 10^{-5}$ \\
$(3^3,4^2)$ &\vline& $0.419615$&\vline&$0.419642$&\vline&$[0.4194,0.4199]$&\vline&$2.7 \times 10^{-5}$ \\
$(3,4,6,4)$ &\vline&$0.524821$&\vline&$0.524833$&\vline&$[0.5246,0.5251]$&\vline&$1.2 \times 10^{-5}$ \\
$(3^2,4,3,4)$ &\vline&$0.414120$ &\vline& $0.414137$&\vline&$[0.4139,0.4144]$&\vline&$1.7 \times 10^{-5}$ \\
\end{tabular}
\end{center}
\caption{Comparison of bond percolation estimates with numerical results (the kagome result is from Feng, Deng and Bl\"ote \cite{Feng}, $(3,12^2)$ is from Ding, Fu, Guo and Wu \cite{Ding}, and the rest are from Parviainen \cite{Parviainen}) and the Riordan and Walters confidence intervals \cite{Riordan}.}
\label{table:bondthresholds}
\end{table}
\subsection{Extensions}
One question that now arises is whether there is any benefit to considering larger unit cells for any of our predictions. The extended checkerboard (Figure \ref{fig:ext_checkerboard}) provides one opportunity to study this issue. For example, we can see what prediction the extended checkerboard formula makes for the simple checkerboard by setting $p_1=p_2=p, r_1=r_2=r, s_1=s_2=s, t_1=t_2=t$. The resulting threshold can be written in factored form:
\begin{eqnarray}
 & &(1 - p r - p s - r s + p r s - p t - r t + p r t - s t + p s t + r s t) \times \nonumber \\
 & &(1 + p r + p s - r s - p r s - p t + r t - p r t + s t - p s t - r s t + 2 p r s t)=0 \nonumber
\end{eqnarray}
We recognize the first term in brackets as our original prediction for the checkerboard, (\ref{eq:checkerboard}), and the second term can be discarded as it does not have a root in $[0,1]$. Evidently, we get the same prediction for the checkerboard, even if we use a larger unit cell. However, if we consider again the striped square lattice shown in Figure \ref{fig:secsquare}, the story is different. Our extended checkerboard formula makes a prediction for this case as we can get Figure \ref{fig:secsquare} from Figure \ref{fig:ext_checkerboard} by substituting $p_1 \rightarrow s_2, r_1 \rightarrow p_1, s_1 \rightarrow s_1, t_1 \rightarrow p_1, p_2 \rightarrow s_1, r_2 \rightarrow p_2, s_2 \rightarrow s_2, t_2 \rightarrow p_2$. This prediction is
\begin{eqnarray}
1 &-& 2 p_1 p_2 + p_1^2 p_2^2 - s_1^2 - 2 p_1^2 s_1 s_2 - 8 p_1 p_2 s_1 s_2 \nonumber \\
 &+& 8 p_1^2 p_2 s_1 s_2 - 2 p_2^2 s_1 s_2 + 8 p_1 p_2^2 s_1 s_2 \nonumber \\
 &-&6 p_1^2 p_2^2 s_1 s_2 + 2 p_1^2 s_1^2 s_2 + 4 p_1 p_2 s_1^2 s_2 \nonumber \\
 &-&4 p_1^2 p_2 s_1^2 s_2 + 2 p_2^2 s_1^2 s_2 - 4 p_1 p_2^2 s_1^2 s_2 \nonumber \\
 &+&2 p_1^2 p_2^2 s_1^2 s_2 - s_2^2 + 2 p_1^2 s_1 s_2^2 + 4 p_1 p_2 s_1 s_2^2 \nonumber \\
 &-&4 p_1^2 p_2 s_1 s_2^2 + 2 p_2^2 s_1 s_2^2 - 4 p_1 p_2^2 s_1 s_2^2 \nonumber \\
 &+&2 p_1^2 p_2^2 s_1 s_2^2 + s_1^2 s_2^2 - 2 p_1^2 s_1^2 s_2^2 \nonumber \\
 &-&2 p_1 p_2 s_1^2 s_2^2 + 2 p_1^2 p_2 s_1^2 s_2^2 - 2 p_2^2 s_1^2 s_2^2 \nonumber \\
 &+&2 p_1 p_2^2 s_1^2 s_2^2=0 \label{eq:secsquare2}
\end{eqnarray}
and equation (\ref{eq:secsquare}) does not factor out. Now, if we set $s_1=s_2=p_2=0.4$, we find $p_1=0.901308$, which, compared to the previous answer $p_1=0.9$, is much closer to the numerical value of $p_c=0.901263(2)$ \cite{ZiffScullard10}. Clearly some refinement of the estimate is possible by considering larger unit cells, at least in some cases.

\begin{center}
\begin{figure}
 \includegraphics{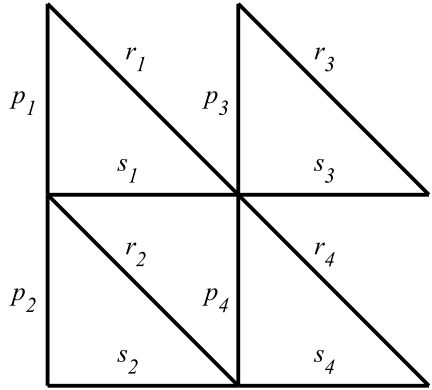} 
\caption{Probability assignments on the extended striped triangle.} \label{fig:extsttr}
\end{figure}
\end{center}

Similarly, we can improve the estimate for the $(3^3,4^2)$ lattice, by considering the triangular lattice with $12$ different probabilities (Figure \ref{fig:extsttr}) rather than $6$ as for the striped triangular problem (Figure \ref{fig:sttr}). We will again list only the constraints used to find $f(p_1,r_1,s_1,p_2,r_2,s_2,p_3,r_3,s_3,p_4,r_4,s_4)$.
\begin{enumerate}
 \item Re-partitioning:
\begin{equation} f(p_1,r_1,s_1,p_2,r_2,s_2,p_3,r_3,s_3,p_4,r_4,s_4)=f(p_2,r_2,s_2,p_1,r_1,s_1,p_4,r_4,s_4,p_3,r_3,s_3) \nonumber
\end{equation}
\item Re-partitioning:
\begin{equation}
 f(p_1,r_1,s_1,p_2,r_2,s_2,p_3,r_3,s_3,p_4,r_4,s_4)=f(p_3,r_3,s_3,p_4,r_4,s_4,p_1,r_1,s_1,p_2,r_2,s_2) \nonumber
\end{equation}
\item Re-partitioning:
\begin{equation}
 f(p_1,r_1,s_1,p_2,r_2,s_2,p_3,r_3,s_3,p_4,r_4,s_4)=f(p_2,r_4,s_3,p_1,r_3,s_4,p_4,r_2,s_1,p_3,r_1,s_2) \nonumber
\end{equation}
\item Reflection:
\begin{equation}
 f(p_1,r_1,s_1,p_2,r_2,s_2,p_3,r_3,s_3,p_4,r_4,s_4)=f(s_4,r_4,p_4,s_2,r_2,p_2,s_3,r_3,p_3,s_1,r_1,p_1)\nonumber
\end{equation}
\item Reflection:
\begin{equation}
 f(p_1,r_1,s_1,p_2,r_2,s_2,p_3,r_3,s_3,p_4,r_4,s_4)=f(s_2,r_1,p_3,s_4,r_3,p_1,s_1,r_2,p_4,s_3,r_4,p_2) \nonumber
\end{equation}
\item Using the $10$-bond bow-tie from Figure (\ref{fig:bowtie10}):
\begin{equation}
 f(p_1,r_1,s_1,p_2,0,s_2,p_3,0,s_3,p_4,r_4,s_4)=\mathrm{EBT}(p_1,s_1,p_3,s_2,r_1,p_4,s_4,p_2,s_3,r_4) \nonumber
\end{equation}
\item 
\begin{equation}
 f(p_1,r_1,s_1,p_2,r_2,s_2,0,r_3,s_3,0,r_4,s_4)=\mathrm{EBT}(s_4,r_3,s_1,r_1,p_1,s_3,r_4,s_2,r_2,p_2) \nonumber
\end{equation}
\item Using (\ref{eq:sttr}):
\begin{eqnarray}
 f(p_1,r_1,s_1,p_2,r_2,s_2,0,0,0,p_4,r_4,s_4)=\mathrm{ST}(s_2,p_2,r_2,s_1,p_1,r_1,s_4,p_4,r_4) \nonumber
\end{eqnarray}
\item 
\begin{eqnarray}
 &f(p_1,r_1,s_1,p_2,r_2,s_2,p_3,r_3,1,p_4,r_4,1) = & \nonumber \\
  &\mathrm{ST}(1 - (1 - p_1)(1 - r_3)(1 - p_3), r_1, s_1, 1-(1-p_4)(1-r_4)(1-p_2),r_2,s_2)& \nonumber
\end{eqnarray}
\item Finally:
\begin{eqnarray}
 &f(p_1,1,s_1,p_2,r_2,s_2,p_3,1,s_3,p_4,r_4,s_4) =& \nonumber \\ 
 &\mathrm{ST}(1 - (1 - p_1)(1 - s_4)(1 - s_1),p_4,r_4,1-(1-s_3)(1-p_3)(1-s_2),p_2,r_2)\ .& \nonumber
\end{eqnarray}
\end{enumerate}
The homogeneous $(3^3,4^2)$ threshold is given by $f(p,0,p,p,p,p,p,0,p,p,p,p)=0$:
\begin{equation}
1 - 4 p^2 - 12 p^3 + 104 p^5 - 193 p^6 + 146 p^7 - 45 p^8 + 2 p^{10}=0
\end{equation}
for which the solution on $[0,1]$ gives $p_c=0.419615...$, the value given in Table \ref{table:bondthresholds}. This is an improvement on the 6-bond estimate of $p_c=0.419308...$ compared to Parviainen's numerical estimate of $p_c=0.4196419(4)$.

Finally, we mention that we also found the 12-bond kagome threshold. However, this makes the same prediction as the 6-bond case, which we know is incorrect. Evidently, to get any improvement on this threshold would require a large number of probabilities. We have seen a few examples where thresholds are unchanged by employing larger unit cells. All those results are either exact or, in the case of the checkerboard and 5-bond bow-tie, thought to be exact. However, we can see here that this property cannot be taken as a test for exactness, as the kagome threshold exhibits some cell-size independence as well.

\section{Discussion}
The results are summarized in Table \ref{table:bondthresholds}, along with the numerical results of Parviainen \cite{Parviainen}, Feng, Deng and Bl\"ote \cite{Feng}, and Ding, Fu, Guo, and Wu \cite{Ding} as well as the confidence intervals of Riordan and Walters \cite{Riordan}. The final two Archimedean graphs, the $(3^4,6)$  and $(4,6,12)$ lattices have large unit cells, containing $15$ and $18$ bonds respectively. This presents a challenge to the method and we have not yet found the polynomials for those systems. However, for the Archimedean solutions we did find, the largest difference between our predicted value and the numerical result is $0.000033$, for the $(4,8^2)$ lattice, and none are ruled out by the Riordan and Walters confidence intervals.

What is not clear, however, is why this linearity argument works as well as it does. If we were to give every bond of a lattice a different probability, then the critical function of this situation would certainly be ``linear'' in this infinite number of probabilities. This is because the threshold could be located (in principle anyway) by enumerating open paths that span increasingly large regions. In any of these paths, a given bond would only be mentioned once, and therefore only the first power of its probability would appear in the solution. Evidently, by extending the inhomogeneous probabilities over ever greater regions, we can approach the exact threshold, and our approximation in which the probabilities cover only a single unit cell of the lattice is the first step in this procedure. However, because the solutions we found are so close to the numerical values, we have effectively shown that convergence to the exact threshold is very fast.

Naturally, we would like to learn how to go beyond these lowest-order results. Although we saw examples of this when we found (\ref{eq:secsquare2}) as a refinement of (\ref{eq:secsquare}) and improved the $(3^3,4^2)$ threshold, finding the higher-order approximations to most problems will likely prove very difficult using the method we have presented here. In our derivations we have been able to avoid the question of what our critical functions actually represent, i.e., perhaps they arise from the application of some comparison of probabilities like condition (\ref{eq:critcond}), so it is conceivable that there is a more computationally straightforward way to find them. Discovering such a specification would be valuable in extending the approximation.

\bibliography{scullard_ziffv5}% Produces the bibliography via BibTeX.

\end{document}